# AWAKE, A Particle-driven Plasma Wakefield Acceleration Experiment

*E. Gschwendtner*
CERN, Geneva, Switzerland

**Abstract**
The Advanced Proton Driven Plasma Wakefield Acceleration Experiment (AWAKE) aims at studying plasma wakefield generation and electron acceleration driven by proton bunches. It is a proof-of-principle R&D experiment at CERN and the world's first proton driven plasma wakefield acceleration experiment. The AWAKE experiment will be installed in the former CNGS facility and uses the 400 GeV/*c* proton beam bunches from the SPS. The first experiments will focus on the self-modulation instability of the long (r.m.s ~12 cm) proton bunch in the plasma. These experiments are planned for the end of 2016. Later, in 2017/2018, low energy (~15 MeV) electrons will be externally injected to sample the wakefields and be accelerated beyond 1 GeV.

## 1   Introduction

The construction of ever larger and costlier accelerator facilities has its limits; accelerating fields of today's RF cavities and microwave technology are limited to about 100 MV/m; hence, several tens of kilometres are required for future linear colliders. New technologies will be needed to push back the energy frontier.

Harnessing wakefields, physicists may be able to produce accelerator gradients hundreds of times higher than those achieved in current radiofrequency cavities. This would allow future colliders to achieve higher energies over shorter distances than is possible today.

With plasma wakefield acceleration, a new energy frontier for particle physics might be realized in an affordable manner; plasma machines could be the vaunted 'table-top' accelerator. Scales of metres rather than kilometres could bring accelerator labs within the reach of any university or industrial laboratory.

For the next generation wakefield accelerator, one could envision so-called 'afterburners', in which existing conventional accelerators, such as the LHC, are used to excite the plasma waves.

Before this new technology can be applied, to fully exploit the high gradients for future accelerators, several key performance parameters need to be benchmarked.

Experimental results have demonstrated the success of the plasma wakefield acceleration and its research: e.g., beam-driven plasma wakefield acceleration experiments [1] performed at Stanford Linear Accelerator Center (SLAC) successfully doubled the energies of some of the electrons in an initial 42 GeV beam in less than 1 m of plasma. Recent results report high-efficiency acceleration of a discrete trailing bunch of electrons containing sufficient charge to extract a substantial amount of energy from the high-gradient, non-linear plasma wakefield accelerator; the core particles gained about 1.6 GeV of energy per particle, with a final energy spread as low as 0.7% and an energy-transfer efficiency from the wake to the bunch that could exceed 30% [2].

## 2 Beam-driven plasma wakefield acceleration experiments: landscape

Table 1 gives a summary of beam-driven plasma wakefield experiments that are either ongoing or under construction.

### 2.1 AWAKE

AWAKE is the first plasma wakefield acceleration experiment worldwide to use a proton beam as a drive beam [3]. The first physics is expected late 2016. AWAKE is a proof-of-principle experiment with the aim of providing a design for a particle physics frontier accelerator within the next decade. AWAKE will be explained in detail in the following sections.

### 2.2 FACET, Stanford Linear Accelerator Center

FACET (Facility for Advanced Accelerator Experimental Tests) is a user facility at the SLAC International Laboratory, providing high energy density electron and positron beams with peak currents of ≈20 kA focused down to $30 \times 30$ μm$^2$ transverse spot size at an energy of 20 GeV [4]. The facility studies the acceleration of high-quality and high-efficiency witness bunch beams, as well as the acceleration of positrons. FACET started in 2012 and hosts more than 150 users and 25 experiments. FACET II has been proposed, to start in 2018.

### 2.3 DESY, Zeuthen

PITZ is the Photo-Injector Test Facility at DESY, Zeuthen. This is a research and development facility [5]; its 20 MeV electron beam is used to study the self-modulation instability (SMI) in a lithium plasma cell.

### 2.4 DESY FLASH Forward

FLASH Forward [6] aims at the advancement of beam-driven plasma wakefield physics towards applications utilizing the expertise and tools provided by the Free Electron Laser facility at DESY. Its goals are the demonstration of capture and controlled release of externally shaped electron beams, the exploration of novel in-plasma beam-generation techniques, and the assessment of those beams for free electron laser gain.

### 2.5 Brookhaven ATF

At Brookhaven National Laboratory [7], experiments are performed to study the quasi-non-linear plasma wakefield acceleration (PWA) regime, driven by multiple bunches and visualization with optical techniques.

### 2.6 SPARC LAB, Frascati

SPARC LAB is a multi-purpose user facility, which includes experiments on laser- and beam-driven plasma wakefield acceleration experiments.

**Table 1:** Beam-driven plasma wakefield experiments

| Facility | Location | Drive beam | Witness beam | Start | End | Goal |
|---|---|---|---|---|---|---|
| AWAKE | CERN, Geneva, Switzerland | 400 GeV protons | Externally injected electron beam (PHIN 15 MeV) | 2016 | 2020, or later | Use for future high energy $e^-/e^+$ collider; Study self-modulation instability; Accelerate externally injected electrons; Demonstrate scalability of acceleration scheme. |
| SLAC-FACET | SLAC, Stanford, USA | 20 GeV electrons and positrons | Two-bunch formed with mask ($e^-/e^+$ and $e^-e^+$ bunches) | 2012 | Sept 2016 | Acceleration of high-quality high-efficiency witness bunch; Acceleration of positrons; FACET II proposal for 2018 operation |
| DESY, Zeuthen | PITZ, DESY, Zeuthen, Germany | 20 MeV electron beam | No witness beam, only drive beam from RF gun | 2015 | ≈2017 | Study self-modulation instability |
| DESY, FLASH Forward | DESY, Hamburg, Germany | X-ray free electron laser type electron beam 1 GeV | Drive and witness in free electron laser bunch, or independent witness bunch (laser wakefield acceleration) | 2016 | 2020+ | Application (mostly) for X-ray free electron laser; Energy-doubling of FLASH-beam energy; Upgrade-stage: use 2 GeV free electron laser drive beam |
| Brookhaven ATF | Brookhaven National Laboratory, USA | 60 MeV electrons | Several bunches, drive and witness formed with mask | Ongoing | | Study quasi-non-linear plasma wakefield acceleration regime; Study plasma wakefield acceleration driven by multiple bunches; Visualization with optical techniques |
| SPARC LAB | Frascati, Italy | 150 MeV electrons | several bunches | Ongoing | | multi-purpose user facility: includes laser- and beam-driven plasmas wakefield experiments. |

## 3 AWAKE: components for a particle-driven plasma wakefield acceleration experiment

### 3.1 Introduction

AWAKE, the advanced proton-driven plasma wakefield acceleration experiment will be installed at CERN in the former CNGS (CERN Neutrinos to Gran Sasso) area and is currently under construction [8]. The first beam for physics experiments is expected by the end of 2016.

AWAKE will use proton bunches, for the first time ever, to drive plasma wakefields [9]. The main physics goals of the experiment are as follows.

— To study the physics of self-modulation of long proton bunches in plasma as a function of beam and plasma parameters. This includes radial modulation and seeding of the instability.

— To probe the longitudinal (accelerating) wakefields with externally injected electrons. This includes measuring their energy spectra for different injection and plasma parameters.

- To study injection dynamics and the production of multigigaelectronvolt electron bunches, either from side injection or from on-axis injection (with two plasma cells). This will include using a plasma density step to maintain the wakefields at the GV/m level over metre distances.
- To develop long, scalable, and uniform plasma cells and develop schemes for the production and acceleration of short bunches of protons for future experiments and accelerators.

In the baseline design of AWAKE at CERN, an LHC-type proton bunch of 400 GeV/$c$ (with an intensity of $3 \times 10^{11}$ protons/bunch) will be extracted from the CERN Super Proton Synchrotron (SPS) and sent along the 750 m long proton beam line towards a plasma cell, which is installed at the upstream end of the CNGS area (see Fig. 1). The proton beam will be focused to $\sigma_{x,y}$ = 200 μm near the entrance of the 10 m long rubidium vapour plasma cell with an adjustable density in the $10^{14}$ to $10^{15}$ electrons/cm$^3$ range.

When the proton bunch, with an r.m.s. bunch length of $\sigma_z$ = 12 cm (0.4 ns), enters the plasma cell, it undergoes the SMI, i.e., the development a long bunch of protons into a series of micro-bunches that resonantly drive large wakefields. The effective length and period of the modulated beam is set by the plasma wavelength (for AWAKE, typically $\lambda_{pe}$ = 1 mm).

A high power (≈4.5 TW) laser pulse, co-propagating and co-axial with the proton beam, will be used to ionize the neutral gas in the plasma cell and also to generate the seed of the proton bunch self-modulation.

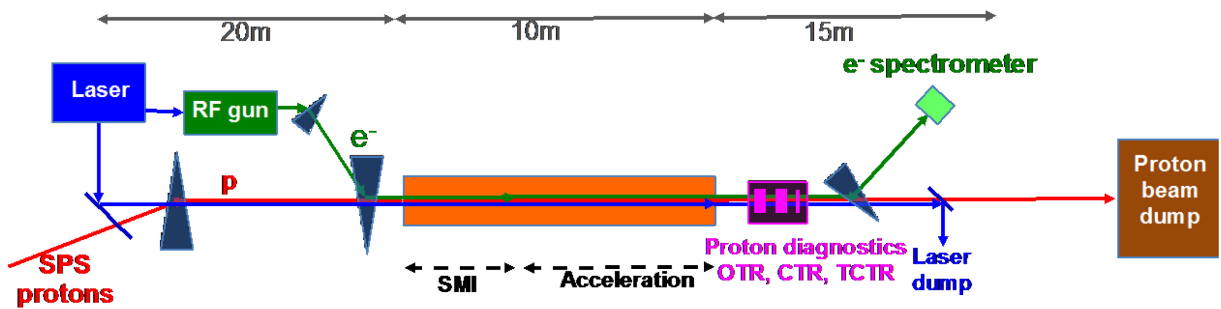

**Fig. 1:** Baseline design of AWAKE

An electron beam of $1.2 \times 10^9$ electrons, which will be injected at 10−20 MeV/$c$ into the plasma cell, serves as a witness beam and will be accelerated in the wake of the proton bunch.

Several diagnostic tools will be installed downstream of the plasma cell to measure the proton bunch self-modulation effects and the accelerated electron bunch properties.

Figure 1 shows the baseline design of the AWAKE experiment. The baseline parameters of the experiment are summarized in Table 2.

In the AWAKE master schedule, the experiment to obtain evidence for the SMI (see Section 3.3) corresponds to Phase 1, and is expected to start by the end of 2016. In Phase 2, AWAKE aims at the first demonstration of proton-driven plasma wakefield acceleration of an electron witness beam; this programme is planned to start by the end of 2017.

**Table 2:** Baseline parameters of the AWAKE experiment

| Parameter | Value | Parameter | Value |
|---|---|---|---|
| **Proton beam** | | **Laser beam to plasma cell** | |
| Momentum | 400 GeV/c | Laser type | Fibre titanium:sapphire |
| Protons/bunch | $3 \times 10^{11}$ | Pulse wavelength | $L_0 = 780$ nm |
| Bunch extraction frequency | 0.03 Hz (ultimate: 0.14 Hz) | Pulse length | 100−120 fs |
| Bunch length | $\sigma_z = 0.4$ ns (12 cm) | Pulse energy | 450 mJ |
| Bunch size at plasma entrance | $\sigma_{x,y} = 200$ μm | Laser power | 4.5 TW |
| Normal emittance (r.m.s.) | 3.5 mm mrad | Focused laser size | $\sigma_{x,y} = 1$ mm |
| Relative energy spread | $\Delta p/p = 0.45\%$ | Energy stability | ±1.5% r.m.s. |
| Beta function | $\beta^*_x = \beta^*_y = 4.9$ m | Repetition rate | 10 Hz |
| Dispersion | $D^*_x = D^*_y = 0$ | | |
| **Electron beam** | | **Plasma source** | |
| Momentum | 16 MeV/c | Plasma type | Laser ionized rubidium vapour |
| Electrons/bunch | $1.2 \times 10^9$ | Plasma density | $7 \times 10^{14}$ cm$^{-3}$ |
| Bunch charge | 0.2 nC | Length | 10 m |
| Bunch length | $\sigma_z = 4$ ps (1.2 mm) | Plasma radius | >1 mm |
| Bunch size at focus | $\sigma_{x,y} = 250$ μm | Skin depth | 0.2 mm |
| Normalized emittance (r.m.s.) | 2 mm mrad | Wavebreaking field, $E_0 = mc\omega_{cp}/e$ | 2.54 GV/m |
| Relative energy spread | $\Delta p/p = 0.5\%$ | | |
| Beta function | $\beta^*_x = \beta^*_y = 0.4$ m | | |
| Dispersion | $D^*_x = D^*_y = 0$ | | |

In the following sections, the different components required for a plasma wakefield experiment will be discussed for the example of the AWAKE project and the choices made in the design explained.

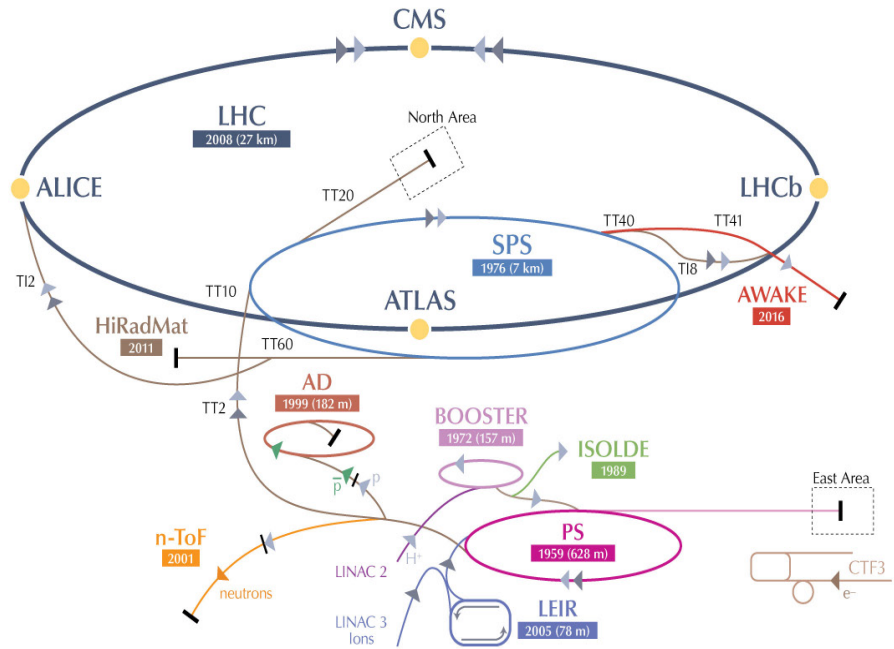

**Fig, 2:** Layout of the CERN accelerator chain

### 3.2 Drive beam

In the baseline design of the AWAKE experiment at CERN, an LHC-type proton bunch of 400 GeV/$c$ will be extracted from the CERN SPS (see Fig. 2).

Simulations for optimization studies were performed [10] with the AWAKE baseline parameters as shown in Table 1. Figure 3 shows the simulated wakefield amplitude of a 400 GeV/$c$ proton beam along a 10 m long plasma cell. With the chosen emittance, the baseline radius of the proton beam of $r = 0.2$ mm is the optimum; wide beams are not dense enough to drive the wave to the limiting field; narrow beams quickly diverge, owing to transverse emittance.

Figure 4 displays the variation of the driver energy at a constant normalized emittance. The nominal beam energy of the SPS at 400 GeV is well suited for the experiment.

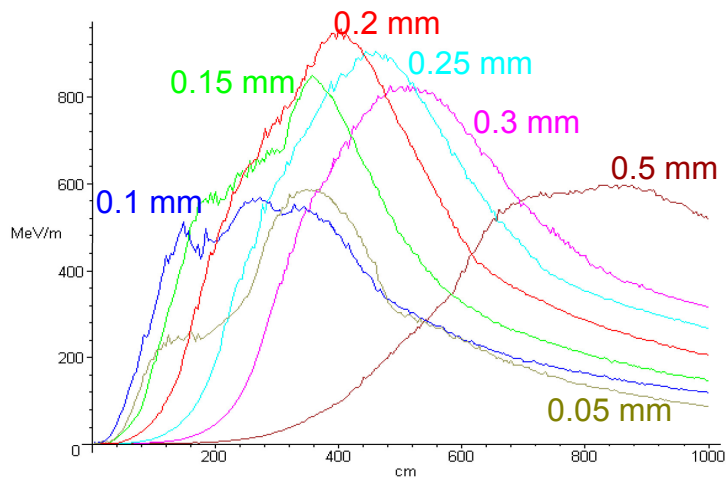

**Fig. 3:** Wakefield driven by 400 GeV proton beam along 10 m long plasma cell with different transverse beam sizes.

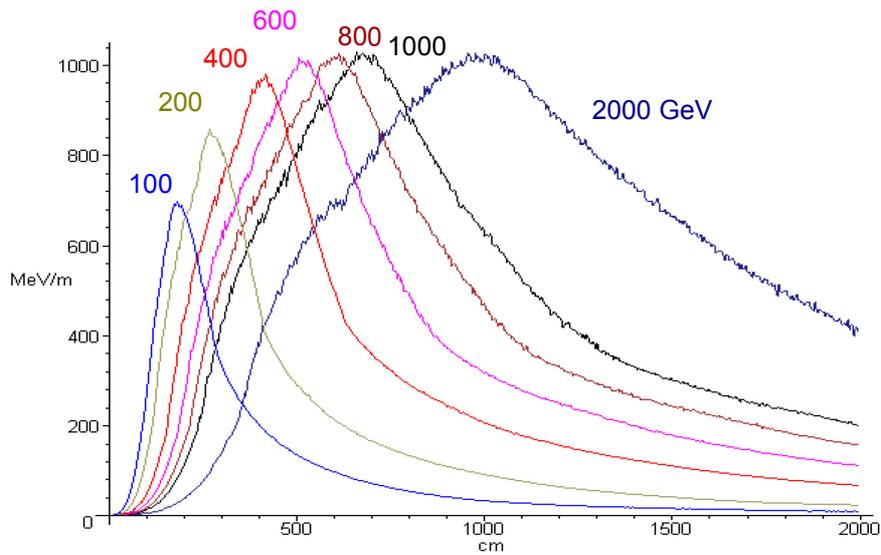

**Fig. 4:** Wakefield driven by proton beam at constant normalized emittance with different energies along plasma cell.

The AWAKE experiment requires short bunches with very high intensity from the SPS, which have not yet been used in operation by any other experiment. The required bunch intensity is above the instability threshold, so these bunches are unstable. If the bunches are unstable, it is difficult to control their parameters (particle distribution and bunch length) from shot to shot. However, they can be stabilized by increasing their longitudinal emittance (and therefore the bunch length) using controlled emittance blow-up.

To obtain smaller bunch lengths, bunches should be rotated in the longitudinal phase space by a quarter of the synchrotron period just before extraction on the SPS flat top (see Fig. 5). This is achieved by reducing and then sharply increasing the RF voltage. This RF manipulation was not used before in the SPS and will be introduced for AWAKE.

The shortest achievable bunch length is determined by the maximum available RF voltage and the smallest possible longitudinal emittance. The maximum voltage in the two RF systems of the SPS is currently 8 MV at 200 MHz and 600 kV at 800 MHz; this will be increased to 12 MV at 200 MHz (by around 2019) and 1.2 MV at 800 MHz (in 2015).

With this method, the SPS proton beam can be optimized to the following parameters: $3 \times 10^{11}$ protons/bunch; normalized transverse emittance, 1.7 mm mrad; r.m.s. bunch length, 9 cm (0.3 ns); peak current, 60 A [11]. However, using conservative estimates, an r.m.s. bunch length of 12 cm (0.4 ns) will be used.

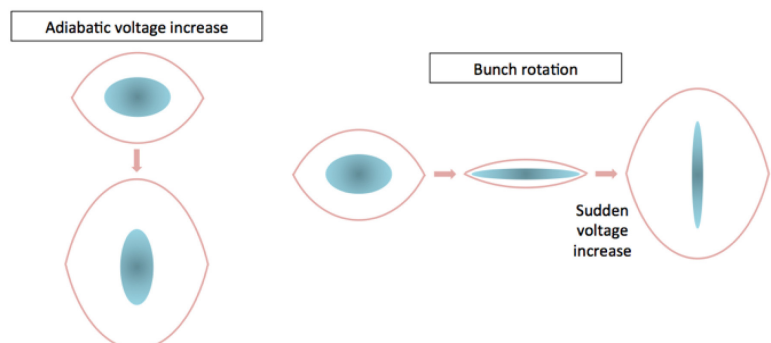

**Fig. 5:** Bunch rotation of SPS beam in longitudinal phase space instead of adiabatic voltage increase

The longitudinal SPS proton beam size of $\sigma_z = 12$ cm is much longer than the plasma wavelength (typically for AWAKE $\lambda_{pe} = 1$ mm). However, the long proton beam can be used directly for plasma wakefield acceleration experiments. This initial approach to proton-driven plasma wakefield acceleration will be through self-modulation of a long bunch of protons into a series of micro-bunches that drive resonantly large wakefields [12].

### 3.3 Self-modulation instability

Plasma wakefields are usually driven by laser pulses or particle bunches approximately one plasma wavelength long: $\sigma_z = \lambda_{pe}$. Here, $\lambda_{pe} = 2\pi c/\omega_{pe}$ is the wavelength of a relativistically moving plasma wave with electron plasma angular frequency $\omega_{pe} = (n_e e^2/\varepsilon_0 \mu_e)^{1/2}$ in a plasma with density $n_e$.

The longitudinal amplitude of the wakefield scales as the wave breaking field [11]:
$$E_{WB} = \mu_e c \omega_{pe}/e \alpha n_e^{1/2}.$$

This scaling therefore favours short pulses or bunches and large plasma densities to reach large amplitude wakefields. These wakefields, for sufficiently intense drivers, have accelerating or decelerating longitudinal components ($E_z \approx E_{WB}$) and transverse focusing or defocusing components with comparable amplitudes. In the linear wakefield regime, these fields vary periodically behind the drive bunch and have a $\pi/2$ phase difference.

Unlike recent plasma wakefield accelerator experiments, which employed short bunches ($\sigma_z < \lambda_{pe}$) to drive intense wakefields, the AWAKE experiment will use much longer proton bunches ($\sigma_z > 100\,\lambda_{pe}$) to generate plasma wakefields.

The AWAKE experiment will thus operate in the so-called self-modulated plasma wakefield accelerator regime [12]. In this regime, the maximum plasma density for wakefield excitation is given by the condition that the plasma return current must flow outside the drive bunch. This condition is satisfied when the bunch transverse size $\sigma_r$ is smaller than the cold plasma collisionless electron skin depth $c/\omega_{pe}$ or when $k_{pe}\sigma_r < 1$ ($k_{pe} = \omega_{pe}/c$). When this condition is not satisfied, the bunch can be subject to the current filamentation instability. The instability breaks the bunch into transverse current filaments [13] and prevents the efficient excitation of plasma wakefields. For $\sigma_r = 200$ μm, setting $k_{pe}\sigma_r = 1$ yields $n_e = 7 \times 10^{14}$ cm$^3$, $\lambda_{pe} = 1.2$ mm and $E_{WB} = 3$ GV/m.

With relativistic bunches, energy gain and loss does not lead to significant dephasing between drive bunch particles over metre-scale plasma lengths:
$$\Delta L \approx \frac{1}{\gamma^2} \frac{\Delta\gamma}{\gamma} L \ll \lambda_{pe}$$

(for particles with energies $\gamma$ and $\gamma \pm \Delta\gamma$ and plasma length $L$). This means that there is no longitudinal bunching.

However, the transverse wakefield components can periodically focus and defocus the particles that typically have non-relativistic transverse velocities,
$$\langle v_\perp \rangle \approx \frac{\varepsilon}{\sigma_0} c \ll c,$$

where $\varepsilon$ is the beam transverse emittance and $\sigma_0$ its waist size [14].

#### 3.3.1 Transverse modulation of a long bunch

When a long and narrow particle bunch travels in a dense plasma; i.e., when $\sigma_z \ll \lambda_{pe}$, it is subject to a transverse two-stream instability or SMI [13]. The low amplitude transverse wakefields driven by the long bunch modulate its radius with wavelength $\sim\lambda_{pe}$. This generates a micro-bunch pattern, as seen in Fig. 6 [15].

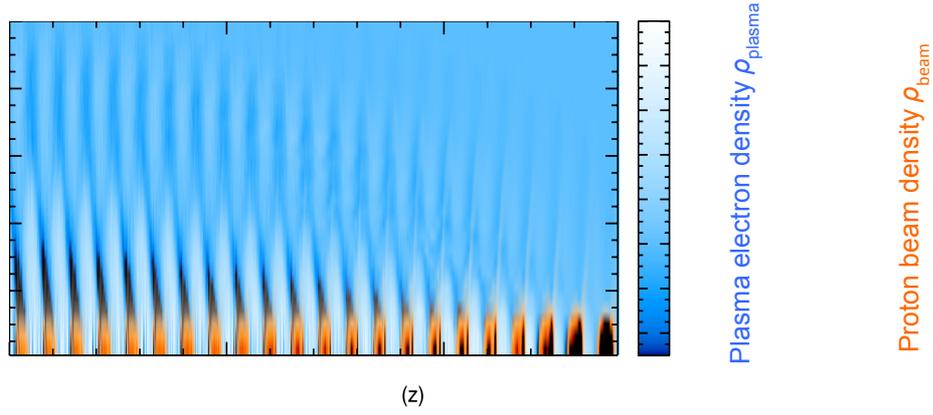

**Fig. 6:** Self-modulated proton bunch resonantly driving plasma wakefields sustained by plasma density perturbation. The plasma electron density is shown increasing from white to blue and the proton density increasing from yellow to dark red.

This periodic bunch density modulation then resonantly drives wakefields to larger amplitudes, thereby providing a feedback mechanism for the SMI to develop. The SMI is a convective instability that grows both along the bunch and along the plasma. If the drive bunch propagates in a uniform density plasma, then the instability destroys the micro-bunches soon after the maximum field is reached [16]. The reason lies in the slow motion of the defocusing field regions with respect to the bunch. This effect causes a strong decrease in the peak accelerating field, as seen in Fig. 7(a).

As the SMI grows, the interplay between bunch radius and wakefield amplitude leads to an effective wakefield phase velocity slower than that of the drive bunch [17, 18], as seen in Fig. 7(b). The figure shows the location of the accelerating and focusing fields along the bunch ($z - ct$) as a function of propagation distance along the plasma, $z$. Once the SMI saturates, these two velocities become equal.

It is possible to avoid the destruction of the micro-bunch structure by a proper step-up in the plasma density (Fig. 7(a)), which modifies the instability growth in such a way that the field motion relative to the bunches stops at the optimal moment.

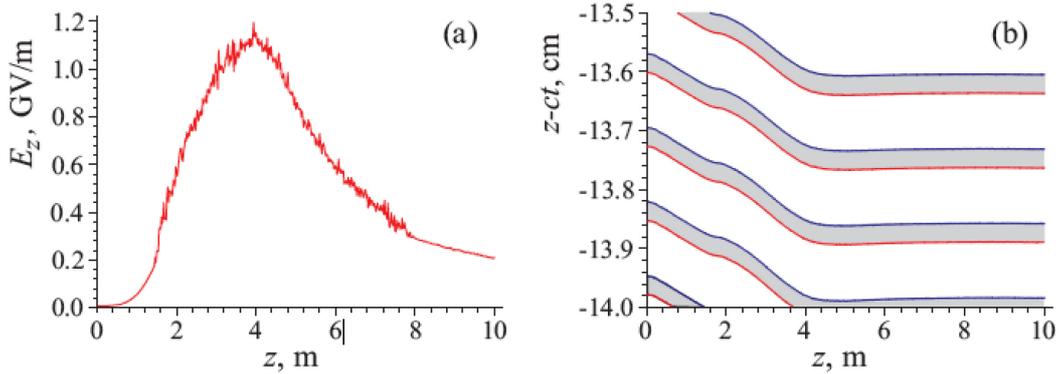

**Fig. 7:** (a) Maximum amplitude of accelerating field $E_z$ excited along the bunch plotted as a function of position along the plasma. (b) Positions along the bunch ($z - ct$) where the wakefields are both accelerating and focusing for witness electrons are shown in grey as a function of propagation along the plasma. This position varies over the first 4 m of propagation and remains at the same $z-ct$ position after that.

### 3.3.2    Seeding of the self-modulation instability

The SMI can, in principle, grow from noise in the plasma and the drive bunch. However, seeding of the instability considerably shortens the plasma length needed for the SMI to reach saturation. Calculations [19] and simulations show that the noise level is very low and that the SMI would not grow to a detectable level over metre-scale plasmas. More importantly, seeding the SMI fixes the phase of the wakefields, a condition necessary to deterministically inject a short witness bunch in the accelerating and focusing phase of the wakefields. The SMI can be seeded by a short laser pulse or particle bunch driving low amplitude wakefields in a preformed plasma ahead of the long drive bunch. A sharp (compared with $\lambda_{pe}$) boundary between the drive bunch and the plasma, such as a cut in the bunch current profile or a relativistic ionization front propagating within the drive bunch, also seeds the SMI [20].

The SMI leads to a radially symmetric modulation of the bunch charge density. However, there is an asymmetric competing instability known as the hose instability [21]. This instability is similar to the beam break-up instability in RF accelerators. It grows from noise in the transverse displacement of the bunch-slice centroids and results in a non-axially symmetric displacement of the bunch along its length. The hose instability has a growth rate comparable to that of the SMI, and the two instabilities directly compete. Simulations indicate that seeding helps the SMI dominate over the hose instability [17, 18].

### 3.4 Plasma cell

The plasma cell for the AWAKE experiment shall fulfil the following requirements.

— Plasma density $n_e$ between $10^{14}$ and $10^{15}$ cm$^{-3}$: Fig. 8 shows the wakefield resulting from a proton beam with AWAKE parameters along a 10 m long plasma cell with different plasma densities [10]. For excessive plasma densities, filamentation and hosing instabilities can occur.

— Allow for seeding.

— Density uniformity $\delta n_e/n_e$ of the order of 0.2 %. The wakefield phase is determined by the plasma density. If the plasma wavelength changes locally, the witness electrons will be defocused. The density must be constant with an accuracy of $\delta_{pe}/4\sigma_z$.

— High-$Z$ gas to avoid background plasma ion motion.

— Gas or vapour easy to ionize.

— Radius $R_p$ larger than $\approx 3$ proton bunch r.m.s. radii.

— Plasma length $L \sim 10$ m.

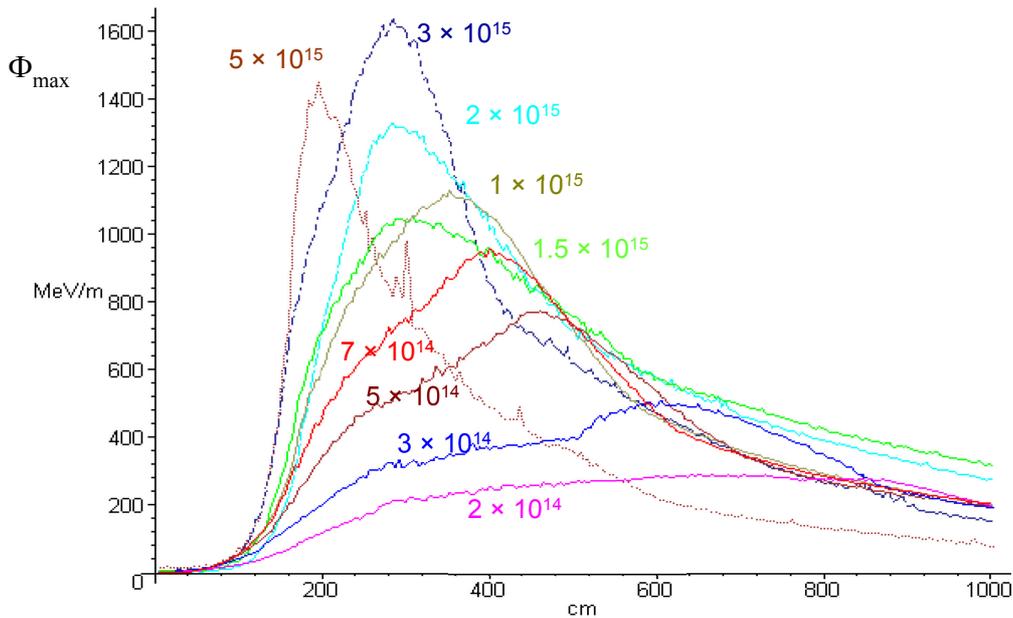

**Fig. 8:** Wakefield along 10 m long plasma cell with different plasma densities $n_e$

Several options exist for plasma cells: discharge plasma sources, helicon sources and metal vapour sources.

AWAKE performs developments using the discharge plasma [22] and helicon sources [23] as these are scalable in length.

In the first phases, and with these requirements, AWAKE will use a rubidium vapour source, 10 m long and 4 cm in diameter [24]. The density uniformity is achieved by controlling the tube containing the vapour to ±0.5 °C. This is achieved by circulating synthetic oil inside a thermal insulation around the tube containing the rubidium vapour. The oil temperature can be stabilized to ±0.01 °C. Figure 9 shows the 10 m long prototype of the AWAKE plasma cell installed in a test area at CERN.

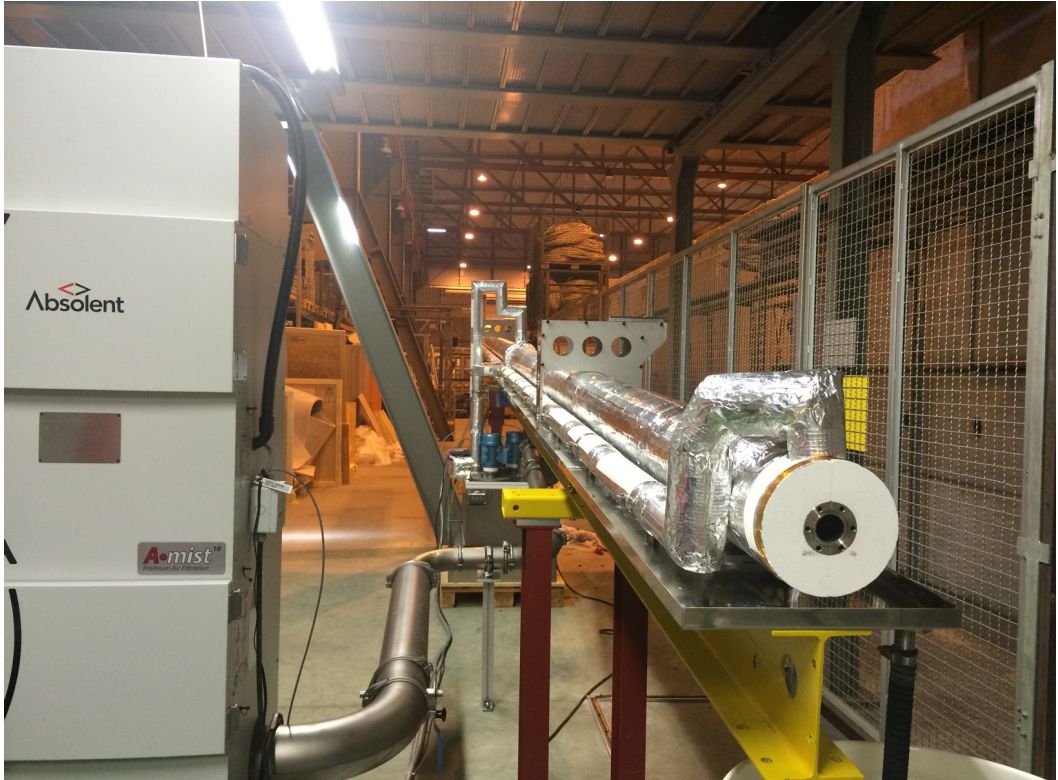

**Fig. 9:** Prototype of 10 m long rubidium vapour plasma cell for AWAKE

A threshold ionization process for the first Rb electron is used to turn the uniform neutral density into a uniform plasma density. The ionization potential is very low, $\Phi_{Rb}$ = 4.177 eV; as is the intensity threshold for over the barrier ionization (OBI), $I_{Ioniz} \approx 1.7 \times 10^{12}$ W/cm$^2$.

### 3.5 Laser beam

The plasma creation method uses a short laser pulse and has many advantages, since it also serves as ionization front seeding method.

A titanium:sapphire laser pulse with 30−100 fs pulse length, $\lambda_0$ = 800 ns and energy of 20−40 mJ provides the energy necessary to ionize the atoms in the plasma and the intensity to ionize along the 10 m long plasma cell [25].

The laser beam moves co-linearly with the proton beam, creating the ionization front, which acts as if the proton beam is sharply cut and in turn directly seeds the SMI.

Figure 10 shows the distribution of the beams after propagating 4 m in the plasma. Protons are blue, electrons are red and the laser pulse is the pink line at $z−ct$ = 0. The laser pulse seeds the SMI for the proton bunch: only the protons that are behind the laser pulse are affected.

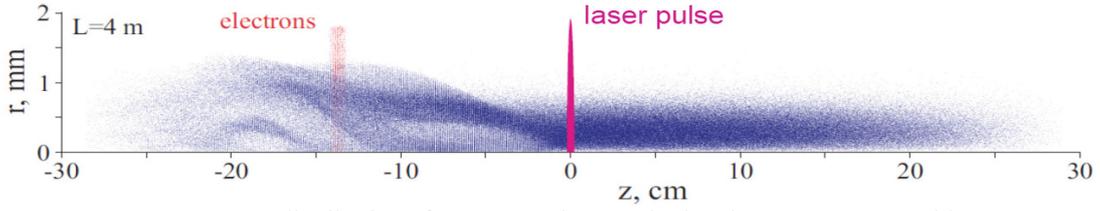

**Fig. 10:** Beam distribution after propagating 4 m in the plasma. Protons are blue

### 3.6 Witness beam

The electron beam is externally injected into the plasma wakefield driven by the proton beam. The considerations for the electron beam specifications are that the electrons must be trapped in the grey area shown in Fig. 7b, where the wakefields are both accelerating and focusing. The optimal electron energy is 10−20 MeV, which corresponds to the wakefield phase velocity at the self-modulation stage.

Table 2 shows the electron beam parameters fitting to the requirements that will be used in AWAKE for the first phase of the experiment. Simulations show a trapping efficiency of the order of ~10% when injecting electrons on-axis at the beginning of the plasma cell [26]. With these parameters and after passing through the 10 m long plasma cell, the expected average energy gain of the electrons will be 1.3 GeV.

Initially, the injected electron bunch will be at least one plasma period long, to avoid precise timing of a much shorter electron bunch with respect to the wakefields [27]. However, in this scheme, the trapping efficiency of electrons is very sensitive to the length of a possible density ramp at the plasma entrance, whereas that of protons is not [12].

The numbers are rather conservative, also allowing the use of the provided electron source without major modifications. However, at a later stage, the requirements for the bunch charge (i.e., 1 nC) and bunch length (i.e., 0.3 ps) will be much tighter, to optimize the performance of the electron acceleration.

Many injection parameters must be optimized in the experiment:

Figure 11 shows the comparison between two different electron injection delays with respect to the laser pulse; in Fig. 11(a), the electron beam is well caught in the wakefield and is accelerated to 1.2 GeV after 10 m. In Fig. 11(b), the different injection delay causes a only very weak acceleration.

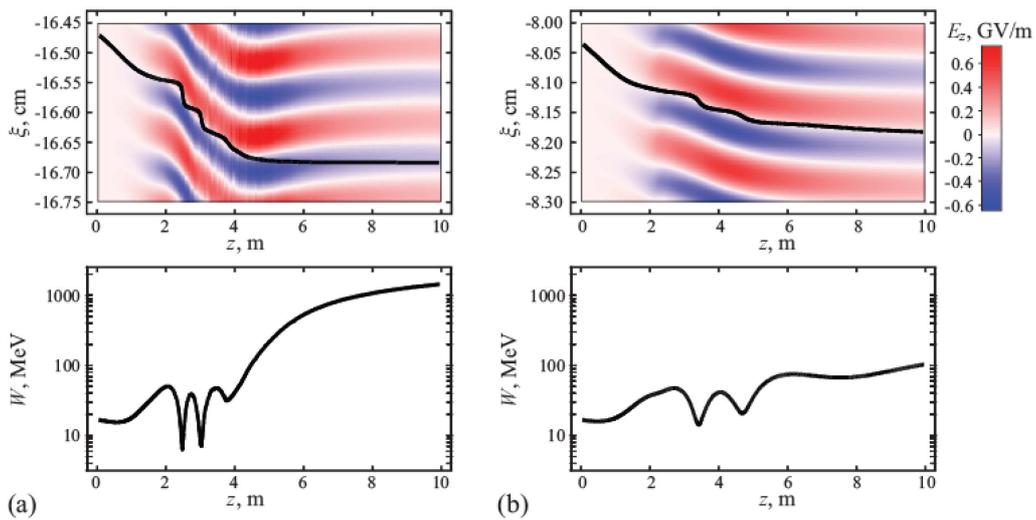

**Fig. 11:** Co-moving coordinate $\xi$ (top) and energy (bottom) versus propagation distance for two typical test electrons injected with different delays with respect to the laser pulse. The top plot also shows the color map of the on-axis electric field $E_z$ in the vicinity of the electron.

### 3.6.1 Electron source

The PHIN Photo-Injector built for CTF3 [28], which will have completed its experimental programme for the Compact Linear Collider at the end of 2015, fulfils the baseline electron beam requirements (10−20 MeV/$c$, $1.2 \times 10^9$ electrons/bunch, $\sigma_z$ =4 ps, $\sigma_{x,y}$ = 250 μm at focus, $\varepsilon_{norm}$ = 2 mm mrad) and will be used as electron source for the AWAKE experiment. The electron source consists of a metal photocathode housed in a 2.5-cell normal conducting RF cavity, a load-lock system for cathode exchange, a 3 GHz booster accelerator, to boost the energy to the required 20 MeV/$c$ and electrical and optical diagnostics (beam position monitors, fast current transformer, Faraday cup, and screens for emittance and spectrometer measurements). In addition, the klystron and modulator system will be recuperated from CTF3, providing 3 GHz RF power.

## 3.7 Diagnostics for the drive beam

The first experiments in AWAKE will be aimed at demonstrating and studying the SMI of the long proton bunch in the dense plasma. Therefore, beam diagnostic tools to measure the radial self-modulation of the bunch density will be developed.

### 3.7.1 Direct SMI measurements

The direct SMI diagnostic tools [29] are based on transformation of the charge distribution information into a radiation distribution using transition radiation. Transition radiation is emitted when a single charged particle or a collection of charged particles cross a boundary between two media with different dielectric constants (e.g., vacuum—metal interface). The radiation is incoherent at wavelengths much shorter than the characteristic charge distribution scale (length or radius) and in the visible range is called optical transition radiation. It is coherent at wavelengths much longer than the characteristic charge distribution scale and is called coherent transition radiation or transverse coherent transition radiation.

With optical transition radiation, the bunch spatial or temporal characteristics are contained in the light intensity and can be measured using, e.g., a streak camera. The best streak cameras have a single pulse temporal resolution of ~1 ps down to ~200 fs. The period of the modulation (expected to be that of the plasma wave) can, in principle, be measured up to frequencies of ~300 GHz. For the AWAKE plasma densities, the plasma and modulation frequency range is 100−300 GHz.

The modulated proton bunch also emits coherent transition radiation, whose spectrum reflects the bunch longitudinal and transverse structure (radial modulation with longitudinal period approximately that of the plasma wave). Therefore, coherent transition radiation emission at the plasma frequency from the radially modulated proton bunch can, in principle, be detected. The long proton bunch radiation can be filtered out by a section of waveguide in cut-off (high-pass filter) and the high-frequency time evolution of the radiation detected using a fast Schottky diode (∼200 ps rise time). In particular, coherent transition radiation emission at high frequencies should be correlated with the ionizing laser position within the proton bunch, i.e., with the SMI seed. The modulation frequency can, in principle, be determined using a heterodyne measurement system, mixing the RF modulation signal of unknown frequency with a known local oscillator frequency in a crystal that generates the difference or intermediate frequency. By choosing appropriate frequencies and a suitable crystal, one can bring the intermediate frequency within the measurement range.

### 3.7.2 Indirect SMI measurements

Another method is to detect the SMI effect by measuring the angular divergence of the proton beam caused by the SMI in the plasma cell, which is of the order of ~1 mrad (M. Turner and A. Petrenko, private communication). For this purpose, the bunch profile is measured at two different scintillator screens installed downstream of the plasma cell and at a distance of ≈8 m with a resolution of ~0.1 mm.

Figure 12(a) shows the energy deposition of the proton beam in the first scintillator screen downstream of the plasma cell; Fig. 12(b) shows the energy deposition in the second scintillator screen, which is ≈8 m downstream of the first one. With this method the saturation point of the SMI inside the plasma cell can also be measured, at a 2% level.

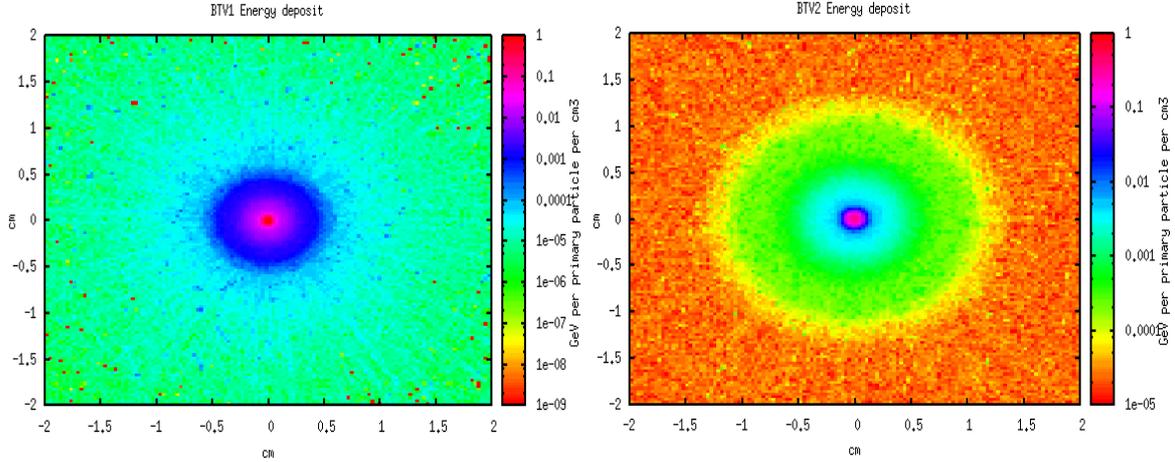

**Fig. 12:** Energy deposition in GeV/cm$^3$ per primary proton particle of the self-modulated proton beam downstream of the plasma cell. (a) Scintillator 1 m downstream of the plasma cell. (b) Scintillator 9 m downstream of the plasma cell.

### 3.7.3   *Witness beam diagnostics*

The purpose of the experiment is to accelerate the externally injected electrons. A magnetic spectrometer was developed with a relatively large energy acceptance, from a few hundred megaelectronvolts to a few gigaelectronvolts [30]. The electron spectrometer system consists of a C-shaped magnet providing a 1.5 T field to separate the electrons from the proton beam and disperse them in energy. The spectrometer system includes a quadrupole doublet in a point-to-point imaging configuration, to focus the beam exiting the plasma onto the spectrometer screen and increase the energy resolution. The electrons impinge on a scintillating screen (the baseline is gadolinium oxysulfide) and an optical line will transport the screen light to a CCD camera. Calculations show that a percentage-level energy resolution can be achieved with a signal-to-noise ratio greater than 1000:1.

## 4   The AWAKE facility—putting the pieces together

The AWAKE experiment will be integrated in an existing underground area, which previously housed the CNGS facility [31]. The different components described in Section 3 must be put together and connected with beam lines.

Figure 13 shows the integration of the AWAKE experiment in the experimental area.

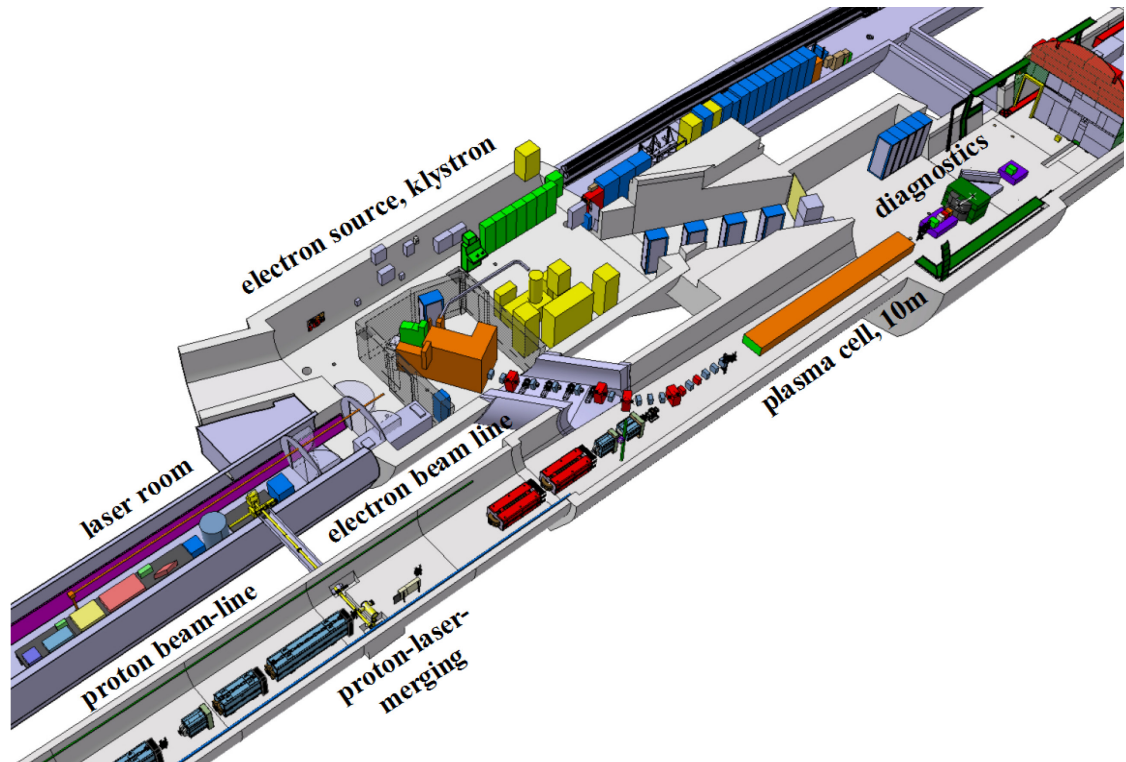

**Fig. 13:** Integration of the AWAKE experiment in the experimental area

### 4.1 Proton beam line

The proton beam is extracted from the CERN SPS and sent along the 850 m long proton beam line towards the plasma cell (see Fig. 2). The existing proton beam line is modified only in its last 80 m or so, i.e., in the matching section and in the final focusing part, to comply with the AWAKE requirements. The laser beam delivered from the laser source is merged with the proton beam at a distance of ≈20 m from the plasma cell by adding a proton beam chicane to integrate the laser mirror. An offset of 19.9 mm exists between the proton and the laser beam axis, enough clearance to avoid intercepting protons and inducing losses [32]. The proton and laser beams are made co-axial over the full length of the plasma cell, in particular the $3\sigma$ proton beam envelope (≈0.6 mm) must be contained in the $1\sigma$ laser spot size (≈1 mm). A pointing precision of 100 µm is required at the cell entrance, resulting in a maximum angular error of 15 µrad for the proton beam line. To achieve this, the ripple in the current of the main dipole power converter is kept below $5 \times 10^{-4}$. A 0.2 mm thick aluminium window will be installed in the proton beam line to separate the SPS and the AWAKE vacuums. Modifications and new beam diagnostics are required at the end of the proton beam line; the electronics of the beam position monitors are modified to allow for single bunch measurements. Two BTVs, located about 1 m upstream and downstream of the plasma cell, will perform profile and position measurements and will also be used during set-up to align the proton and laser beams. The synchronization between the two beams will be adjusted with a streak camera connected to a BTV, located ~2 m upstream of the plasma cell.

### 4.2 Laser beam line

A gallery previously used for CNGS equipment storage is modified to become a dust-free (over-pressurized) and temperature-stabilized area to house the laser system. Owing to the small size of the gallery, the integration of the laser, pulse compressor, laser beam transport optics, laser tables, racks, and clean-room features, such as a special cooling and ventilation system, and a double door access system, is very challenging, but has been successfully achieved [33]. Two laser beams are to be transported from the laser room. The first laser beam is used for plasma production and seeding of the

SMI in the proton bunch. For that aim a 4 m long laser core (50 cm in diameter) was drilled between the laser room and the proton beam line tuning. As the peak power of the 100 fs compressed laser pulse is very high, the laser beam must be injected into the vacuum system of the SPS proton beam before laser pulse compression. Therefore, the vacuum laser beam line connecting the laser with the proton merging area has to fulfil the requirements of the SPS vacuum system ($10^{-7}$ mbar), resulting in a demanding design for the laser line optics and vacuum system.

The second laser beam is required for the generation of the electron beam and has to be transported to the electron source.

### 4.3 Electron beam line

A new 15 m long electron beam line has been designed to transport the electron beam from the RF gun, across a newly built tunnel (7 m long, 2.5 m wide), towards the proton beam line to be injected into the plasma cell on the same axis as the proton and laser beam. To comply with experimental requirements, the electron beam optics must provide a flexible design so that the focal point can be varied by up to 6 m inside the plasma cell [34]. The proton and the electron beams share the last ≈5 m of the line upstream of the plasma cell. The technical parameters for the new magnets (4 dipoles, 11 quadrupoles, and 11 correctors) have been specified. In addition, specifications were defined for the electron beam line diagnostics; e.g., the beam position monitors should be able simultaneously to measure the position of electrons in the presence of the AWAKE proton beam. A peak-to-peak resolution of 50 μm is required. The electron beam shall be synchronized with the proton and laser beam at a <1 ps level, and will be measured with a streak camera linked to a BTV. To meet the performance criteria of the light transmission system, a straight line between the beam screen and the camera was required, which implied the excavation of a ≈4 m long 15 cm diameter core between the proton line and the electron source area, where the streak camera will be installed.

### 4.4 Beam synchronization

Synchronization between the laser pulse and the electron beam at the level of a few tens of femtoseconds (a fraction of the plasma period of ≈4 ps) is required for the deterministic injection of the witness bunch into the plasma wakefields. This is achieved by driving the RF gun of the electron source with a laser pulse derived from the same laser system used for plasma ionization and seeding. The synchronization between the proton and the laser beam must be better than 100 ps, i.e., better than the r.m.s. proton bunch length of ≈400 ps. The laser mode-locker ($f_{ML}$ = 88.17 MHz) cannot follow the relative revolution frequency change through the SPS acceleration cycle, and therefore the proton bunch in the SPS must be synchronized to the AWAKE reference prior to extraction.

A low phase-noise RF source at 3 GHz is made the master, to which the 34th harmonic of the laser oscillator (at $f_{ML}$) is locked [35]. Lower frequencies for fast triggers (176.3 MHz), SPS synchronization (8.68 kHz) and laser or electron beam repetition rate (10 Hz) are derived by integer division from the master oscillator. Only the 200 MHz RF source for the synchronization with the SPS requires a fractional divider.

A system is foreseen that allows the exchange of synchronization signals on ≈3 km long fibres between the AWAKE facility and the SPS RF Faraday cage; it has been verified that the jitter and drift of the signal transmission do not exceed the picosecond range, slow variations will be compensated by a feedback loop.

## 5    Summary

The AWAKE experiment will, for the first time, use a proton bunch to drive wakefields in a plasma. The first experiments, scheduled to begin in late 2016, will aim at studying the development and seeding of the SMI of the proton bunch in a ≈10 m long, $10^{14}$ to $10^{15}$ cm$^{-3}$ electron density plasma. Later experiments, scheduled for 2018, will study the injection of a RF-gun-produced long electron bunch in the wakefields and its acceleration to gigaelectronvolt energies. Future experiments will use two plasma sources and an ultra-short electron bunch to address accelerator related issues [29]. Long-term prospects include using a short proton bunch to drive wakefields without resort to SMI with the possibility of accelerating electrons to very high energies, possibly at the energy frontier.